\documentclass{PoS}
\usepackage{multirow}
\usepackage{braket}
\usepackage{graphicx}

\title{Model approach to the sign problem on lattice QCD with theta vacuum}

\ShortTitle{Model approach to the sign problem on lattice QCD with theta vacuum}

\author{\speaker{Takahiro Sasaki}\\
        Department of Physics, Graduate School of Sciences, Kyushu University, Fukuoka 812-8581, Japan\\
        E-mail: \email{sasaki@phys.kyushu-u.ac.jp}}

\author{Junichi Takahashi\\
        Department of Physics, Graduate School of Sciences, Kyushu University, Fukuoka 812-8581, Japan\\
        E-mail: \email{takahashi@phys.kyushu-u.ac.jp}}

\author{Yuji Sakai\\
        Quantum Hadron Physics Laboratory, RIKEN Nishina Center, Saitama 351-0198, Japan\\
        E-mail: \email{ysakai@riken.ac.jp}}

\author{Hiroaki Kouno\\
        Department of Physics, Saga University, Saga 840-8502, Japan\\
        E-mail: \email{kounoh@cc.saga-u.ac.jp}}

\author{Masanobu Yahiro\\
        Department of Physics, Graduate School of Sciences, Kyushu University, Fukuoka 812-8581, Japan\\
        E-mail: \email{yahiro@phys.kyushu-u.ac.jp}}

\abstract{
We propose a practical way of circumventing the sign problem in lattice QCD simulations with a theta-vacuum term.
This method is the reweighting method for the QCD Lagrangian after the $SU_A(3)\otimes U_A(1)$ transformation.
In the Lagrangian, the $P$-odd mass term as a cause of the sign problem is minimized.
Additionally, we investigate theta-vacuum effects on the QCD phase diagram for the realistic 2+1 flavor system, using the three-flavor Polyakov-extended Nambu-Jona-Lasinio (PNJL) model and the entanglement PNJL model as an extension of the PNJL model. 
The theta-vacuum effects make the chiral transition sharper.
We finally investigate theta dependence of the transition temperature and compare with the result of the pure gauge lattice simulation with imaginary theta parameter.
}

\FullConference{
The 30 International Symposium on Lattice Field Theory - Lattice 2012,\\
June 24-29, 2012\\
Cairns, Australia
}

\begin{document}
\section{Introduction}
\indent
The existence of instanton solution requires the QCD Lagrangian with the theta vacuum:
\begin{equation}
\mathcal{L}
=
\sum_f
\bar{q}_f(\gamma_{\nu}D_{\nu}+m_f)q_f
+
\frac{1}{4g^2}F^a_{\mu\nu}F^a_{\mu\nu}
-
i\theta\frac{1}{64\pi^2}\varepsilon_{\mu\nu\sigma\rho}F^a_{\mu\nu}F^a_{\sigma\rho},
\label{L_QCD}
\end{equation}
\noindent
in Euclidean spacetime.
Though the angle $\theta$ can take any arbitrary value theoretically, 
experimental measurements of neutron dipole moment give the upper limit of theta, $|\theta |<10^{-9}$\cite{Baker,Kawarabayashi}.
Why should $\theta$ be so small?
This long-standing puzzle is called the strong $CP$ problem.
\\
\indent
Since the upper limit is determined only at zero temperature, the behavior 
is nontrivial for finite temperature. Hence 
the first-principle lattice simulation is needed, but 
it has the sign problem for finite $\theta$, 
since the action is complex there. 
For this reason the lattice simulation is only performed 
by Taylor expansion around $\theta =0$ or analytic continuation 
from the imaginary $\theta$ region
where the action is real~\cite{Negro}.
On the other hand, after making $U_A(1)$ transformation
\begin{equation}
q=e^{i\gamma_5\frac{\theta}{2N_f}}q',
\end{equation}
\noindent
$\theta$ dependence appears only through the mass term 
\begin{eqnarray}
m_f(\theta )
=
m_f\cos (\theta /N_f)+m_fi\gamma_5\sin (\theta /N_f),
\end{eqnarray}
in the transformed Lagrangian
\begin{eqnarray}
\mathcal{L}
=
\sum_f
\bar{q}_f(\gamma_{\nu}D_{\nu}+m_f(\theta))q_f
+
\frac{1}{4g^2}F^a_{\mu\nu}F^a_{\mu\nu}.
\end{eqnarray}
\noindent
The $P$-odd mass term including $i\gamma_5$ 
makes the fermion determinant complex.
\\
\indent
We propose the following approach in order to circumvent this sign problem\cite{Sasaki-theta}.
Performing $SU_A(3)\otimes U_A(1)$ transformation, 
\begin{equation}
q_u=e^{i\gamma_5\frac{\theta}{4}}q_u^{~\prime}~,~~q_d=e^{i\gamma_5\frac{\theta}{4}}q_d^{~\prime}~,~~q_s=q_s^{~\prime}, 
\label{31_trans}
\end{equation}
\noindent
one can find that Lagrangian (\ref{L_QCD}) becomes the following form
\begin{eqnarray}
\mathcal{L}
&=&
\sum_{l=u,d}
\bar{q}_l^{~\prime}\mathcal{M}_l(\theta )q_l^{~\prime}
+
\bar{q}_s^{~\prime}\mathcal{M}_sq_s^{~\prime}
+
\frac{1}{4g^2}F^a_{\mu\nu}F^a_{\mu\nu},
\\
\mathcal{M}_l(\theta )
&=&
\gamma_{\nu}D_{\nu}+m_l\cos (\theta /2)+m_li\gamma_5\sin (\theta /2),
\\
\mathcal{M}_s
&=&
\gamma_{\nu}D_{\nu}+m_s.
\end{eqnarray}
\noindent
Here $\theta$ dependence appears only in the light-quark-mass term and 
the sign problem is induced by the $P$-odd term.
However, the scale of this $P$-odd term is much smaller than $\Lambda_{\rm QCD}$ and hence this term is expected to be negligible.
If the $P$-odd mass term is neglected in the reference theory of 
the reweighting method, the expectation value of operator $\mathcal{O}$ 
is obtained as 
\begin{eqnarray}
\braket{\mathcal{O}}
&=&
\int\mathcal{D}A~\mathcal{O}^{~\prime}({\rm det}\mathcal{M}_l^{~\prime}(\theta ))^2{\rm det}\mathcal{M}_se^{-S_g},
\label{reweight}
\\
\mathcal{M}_l^{~\prime}(\theta )
&=&
\gamma_{\nu}D_{\nu}+m_l\cos (\theta /2),
\\
\mathcal{O}^{~\prime}
&=&
\mathcal{O}
\frac{({\rm det}\mathcal{M}_l(\theta ))^2}{({\rm det}\mathcal{M}'_l(\theta ))^2},
\end{eqnarray}
\noindent
with the gluon part $S_g$ of the QCD action.
If this reference system is good, we can expect $\mathcal{O}^{~\prime}\approx\mathcal{O}$ and calculate $\braket{\mathcal{O}}$ with good accuracy.
\\
\indent
In order to justify our proposal, we use effective models and investigate vacuum condensates.
First, we compare two results with and without the $P$-odd mass term and examine the effect of the neglect.
Secondary, we investigate the phase structure in the $\theta$-$T$-$\mu$ space and $\theta$ dependence of transition temperatures.
This work is mainly based on the Ref. \cite{Sasaki-theta}.
\section{Model setting}
\indent
We use the three-flavor Polyakov-loop extended Nambu-Jona-Lasinio (PNJL) model
\begin{eqnarray}
\mathcal{L}
&=&
\bar{q}_f(\gamma_{\nu}D_{\nu}+\hat{m}_f-\gamma_4\hat{\mu})q_f
-G_{\rm S}\sum^8_{a=0}\left[
(\bar{q}\lambda_aq)^2+(\bar{q}i\gamma_5\lambda_aq)^2
\right]
\nonumber
\\
&&
+G_{\rm D}\left[
{\rm det}\bar{q}_f(1-\gamma_5)q_{f'}
+
{\rm det}\bar{q}_f(1+\gamma_5)q_{f'}
\right]
+
\mathcal{U}(T,\Phi [A],\Phi^*[A]),
\end{eqnarray}
\noindent
where $D_{\nu}=\partial_{\nu}-i\delta_{\nu 4}A_4$ and the Gell-Mann matrices $\lambda_a$ act on the flavor space. 
The three-flavor quark fields $q=(q_u,q_d,q_s)$ have masses $\hat{m}_f={\rm diag}(m_u,m_d,m_s)$, 
and the chemical potential matrix $\hat{\mu}={\rm diag}(\mu , \mu , \mu )$ is defined with the quark-number chemical potential $\mu$. 
Parameters $G_{\rm S}$ and $G_{\rm D}$ denote coupling constants of the scalar-type four-quark and 
the Kobayashi-Maskawa-'t Hooft (KMT) determinant interaction \cite{tHooft,Kobayashi-Maskawa}, respectively, 
where the determinant runs in the flavor space.
The KMT determinant interaction breaks the $U_A(1)$ symmetry explicitly. 
We use the Polyakov potential $\mathcal{U}$ of Ref. \cite{Roessner}:
\begin{eqnarray}
\frac{\mathcal{U}(T,\Phi [A],\Phi^*[A])}{T^4}
&=&
\left[
-\frac{a(T)}{2}\Phi^*\Phi+b(T)\ln \left(
1-6\Phi^*\Phi+4(\Phi^3+\Phi^{*3})-3(\Phi^*\Phi )^2
\right)
\right],
\\
a(T)&=&a_0+a_1\left(\frac{T_0}{T}\right)+a_2\left(\frac{T_0}{T}\right)^2
~,~~
b(T)=b_3\left(\frac{T_0}{T}\right)^3,
\end{eqnarray}
as a function of the traced Polyakov-loop $\Phi$.
\\
\indent
The four-quark vertex $G_{\rm S}$ is originated in a one-gluon exchange between quarks and its higher-order diagrams hence the $G_{\rm S}$ can depend on $\Phi$.
We simply assume the following form for $G_{\rm S}$ \cite{Sakai-epnjl},
\begin{equation}
G_{\rm S}
\rightarrow
G_{\rm S}(\Phi )
=
G_{\rm S}\left[
1-\alpha_1\Phi^*\Phi
-\alpha_2 (\Phi^3+\Phi^{*3})
\right]
\end{equation}
\noindent
which preserves the chiral symmetry, the charge conjugation ($C$) symmetry and the extended $\mathbf{Z}_3$ symmetry \cite{Sakai-Z3}.
The effective vertex $G_{\rm S}$ is called the entanglement vertex and the PNJL model with this vertex is the EPNJL model. 
It is expected that dependence of $G_{\rm S}$ will be determined in future by the accurate method such as the exact renormalization group method\cite{Kondo,Braun,Wetterich}.
The parameters $\alpha_1$ and $\alpha_2$ are fitted to $(\alpha_1,\alpha_2)=(0.25,0.1)$, to reproduce the result of degenerate three-flavor LQCD with imaginary chemical potential\cite{Sasaki-Columbia}.
\\
\indent
The EPNJL model has good consistency with lattice results.
For the transition temperature, the PNJL model is good for the deconfinement transition but overestimates the lattice data for the chiral transition.
However the EPNJL model well reproduces both of the lattice data.
Additionally, properties of QCD in the pure imaginary potential region are important.
In this region, it has a periodicity in ${\rm Im}(\mu )/T$ called the Roberge-Weiss (RW) periodicity\cite{RW}.
The PNJL and EPNJL models also have the periodicity.
At boundaries of the RW period, there is a critical point called the RW endpoint.
Recently, quark-mass dependence of this endpoint was investigated by lattice simulations\cite{Forcrand,DElia}.
The EPNJL model is successful in reproducing the LQCD result, while the PNJL model cannot reproduce this property\cite{Sasaki-Columbia}.
\\
\indent
$\theta$ dependence of the EPNJL model is introduced through the KMT interaction:
\begin{eqnarray}
\mathcal{L}
&=&
\bar{q}_f(\gamma_{\nu}D_{\nu}+\hat{m}_f-\gamma_4\hat{\mu})q_f
-G_{\rm S}(\Phi )\sum^8_{a=0}\left[
(\bar{q}\lambda_aq)^2+(\bar{q}i\gamma_5\lambda_aq)^2
\right]
\nonumber\\
&&+G_{\rm D}\left[
e^{i\theta}
{\rm det}\bar{q}_f(1-\gamma_5)q_{f'}
+
e^{-i\theta}
{\rm det}\bar{q}_f(1+\gamma_5)q_{f'}
\right]
+
\mathcal{U}(T,\Phi [A],\Phi^*[A]).
\end{eqnarray}
\noindent
Performing the chiral transformation (\ref{31_trans}), one can get
\begin{eqnarray}
\mathcal{L}
&=&
\bar{q}_f^{~\prime}(\gamma_{\nu}D_{\nu}+m_f(\theta )-\gamma_4\hat{\mu})q_f^{~\prime}
-G_{\rm S}(\Phi )\sum^8_{a=0}\left[
(\bar{q}^{~\prime}\lambda_aq^{~\prime})^2+(\bar{q}^{~\prime}i\gamma_5\lambda_aq^{~\prime})^2
\right]
\nonumber\\
&&
+G_{\rm D}\left[
{\rm det}\bar{q}_f^{~\prime}(1-\gamma_5)q_{f'}^{~\prime}
+
{\rm det}\bar{q}_f^{~\prime}(1+\gamma_5)q_{f'}^{~\prime}
\right]
+
\mathcal{U}(T,\Phi [A],\Phi^*[A]),
\\
m_l(\theta )
&=&
m_l\cos (\theta /2)+m_li\gamma_5\sin (\theta /2),
\\
m_s(\theta )
&=&
m_s.
\end{eqnarray}
\noindent
Here $\theta$ dependence appears only in the mass term for light quarks.
The $P$-odd term has a much smaller scale than $\Lambda_{\rm QCD}$.
\\
\indent
Since $P$ symmetry is broken at finite $\theta$, we consider $P$-even and $P$-odd condensates,
\begin{eqnarray}
\sigma_f^{~\prime}&\equiv&\braket{\bar{q}_f^{~\prime}q_f^{~\prime}}\\
\eta_f^{~\prime}&\equiv&\braket{\bar{q}_f^{~\prime}i\gamma_5q_f^{~\prime}}
\end{eqnarray}
with $f=l,s$ and assume isospin symmetry ($\sigma_u^{~\prime}=\sigma_d^{~\prime}\equiv\sigma_l^{~\prime},\eta_u^{~\prime}=\eta_d^{~\prime}\equiv\eta_l^{~\prime}$).
The vacuum condensates $X=\sigma_f^{~\prime}, \eta_f^{~\prime}, \Phi$ and $\Phi^*$ are determined by the stationary conditions,
\begin{equation}
\frac{\partial{\Omega}}{\partial X}=0,
\end{equation}
where $\Omega$ is thermodynamic potential calculated with the mean field approximation.
\section{Numerical results}
\begin{figure}[t]
\begin{center}
\includegraphics[height=0.3\textheight,bb= 85 50 240 210,clip]{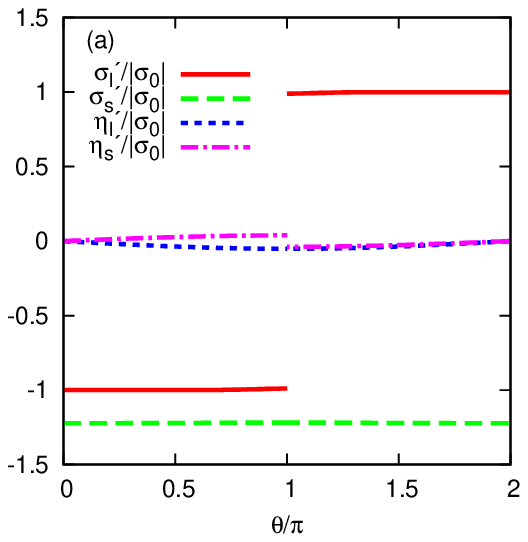}
\includegraphics[height=0.3\textheight,bb= 85 50 240 210,clip]{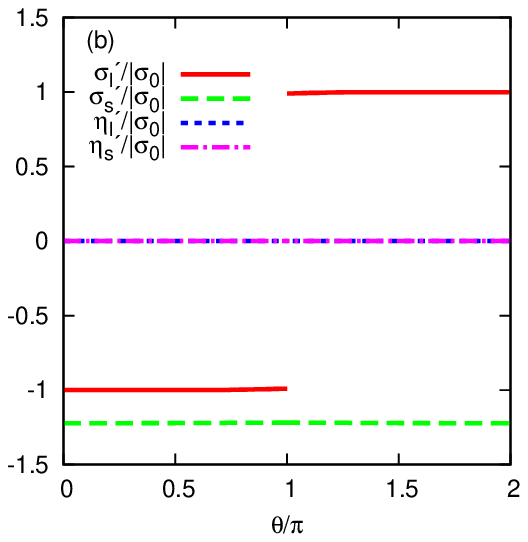}
\caption{
$\theta$ dependence of the order parameters at $T=\mu =0$ in the EPNJL model.
Panel (a) shows a result with the $P$-odd mass and panel (b) corresponds to a result without the $P$-odd mass.
}
\label{fig1}
\end{center}
\end{figure}

\begin{figure}[t]
\begin{center}
\includegraphics[width=0.49\textwidth,bb= 50 50 210 165,clip]{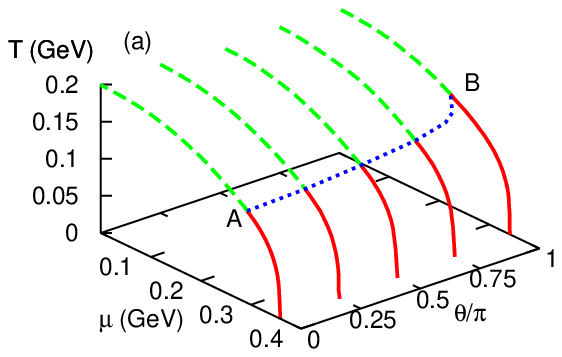}
\includegraphics[width=0.49\textwidth,bb= 50 50 210 165,clip]{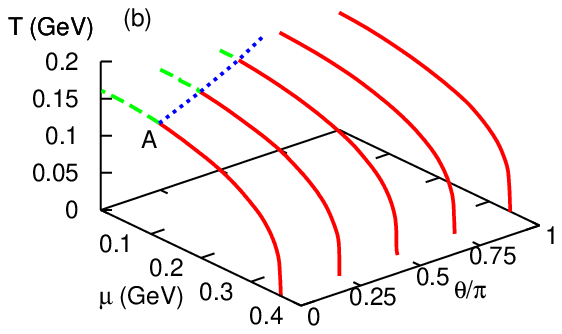}
\caption{
Phase diagram of the chiral transition in the $\mu$-$\theta$-$T$ space.
Panel (a) shows a result of the PNJL model and panel (b) corresponds to a result of the EPNJL model.
}
\label{fig2}
\end{center}
\end{figure}

\begin{figure}[t]
\begin{center}
\includegraphics[height=0.3\textheight,bb= 70 50 250 210,clip]{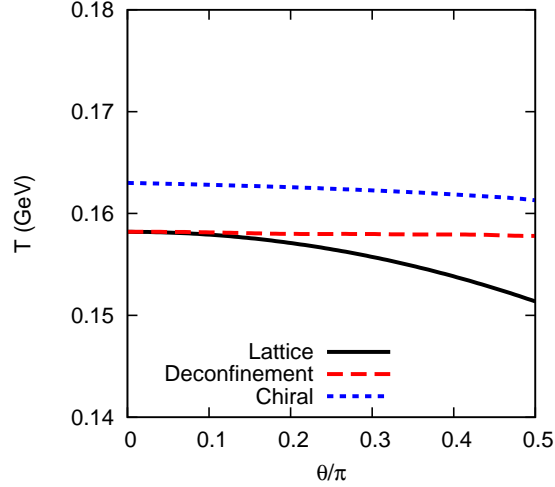}
\caption{
$\theta$ dependence of transition temperatures at $\mu =0$.
The solid line shows the lattice result for the deconfinement transition\cite{Negro}.
Dashed (dotted) lines correspond to result of the EPNJL model for the deconfinement (chiral) transitions.
}
\label{fig3}
\end{center}
\end{figure}

\indent
Figure \ref{fig1} shows $\theta$ dependence of condensates at $T=\mu =0$.
Since $\Phi =0$ at zero temperature, the PNJL and EPNJL models give the same result as each other.
In Fig. \ref{fig1}(a), the $P$-odd mass is taken into account.
Solid and dashed lines show the $P$-even condensates for light and strange quarks, 
while dotted and dot-dashed lines correspond to the $P$-odd condensates.
Since $P$-odd condensates are much smaller than $P$-even condensates, 
$P$-odd mass is expected to be negligible for this case.
In Fig. \ref{fig1}(b), the $P$-odd mass term is neglected.
$P$-odd condensates ($\eta_l{}',\eta_s{}'$) become zero, but $P$-even condensates ($\sigma_l{}',\sigma_s{}'$) are not affected by the neglect.
Therefore it is expected that vacua with and without the $P$-odd mass term are similar to each other and the reweighting method (\ref{reweight}) works well.
\\
%
%
\indent
Figure \ref{fig2} show phase diagrams in the $T$-$\mu$ plane as a function of $\theta$ obtained by (a)the PNJL and (b)the EPNJL model.
Solid lines show the first order chiral transition and dashed lines correspond to the chiral crossover.
Hence, point A is a critical endpoint at $\theta =0$.
The endpoint slightly moves to smaller $\mu$ and higher $T$ when $\theta$ is increased from $0$ to $\pi$.
In the PNJL model, the critical endpoint does not disappear even for $\theta =\pi$.
However, for the EPNJL model, this movement is much faster and there is no critical endpoint at large $\theta$.
This means a possibility that the cosmic evolution is changed at QCD epoch by the first order transition if $\theta$ is large.
\\
%
%
\indent
Figure \ref{fig3} shows $\theta$ dependence of transition temperatures at $\mu =0$.
Dashed and dotted lines show EPNJL model results for the deconfinement and the chiral transition temperature, respectively.
Solid line shows a result of lattice simulations\cite{Negro}:
\begin{eqnarray}
\frac{T_c(\theta )}{T_c(0)}
&=&
1-R_{\theta}\theta^2+O(\theta^4),
\\
R_{\theta}
&=&
0.0175(7).
\end{eqnarray}
\noindent
The coefficient $R_{\theta}$ has been determined by lattice simulations of pure Yang-Mills theory with imaginary $\theta$ parameter, 
and the constant $T_c(0)$ is fixed to that of EPNJL model.
Compared with the lattice result, $\theta$ dependence of model result is much smaller.
This result shows that lattice simulations with dynamical quarks are crucial for theta vacuum effects.
%
%
\section{Summary}
\indent
The QCD Lagrangian with $\theta$ vacuum has the sign problem 
because of the topological term.
This term can be vanished by the chiral transformation (\ref{31_trans}), 
but the transformed Lagrangian has theta dependence in its 
light-quark mass terms.
Using the fact that the $P$-odd mass has a much smaller scale than 
$\Lambda_{QCD}$, we have proposed a way of circumventing 
the sign problem. 
The reweighting method defined by (\ref{reweight}) may allow us to do LQCD calculations and get definite results on dynamics of $\theta$ vacuum.
\\
\indent
Furthermore, we have investigated effects of the theta vacuum on the 
phase diagram for the realistic 2 + 1 flavor system, using the three-flavor PNJL and EPNJL models. 
Particularly in the EPNJL model that is more reliable than the PNJL model, the transition becomes first-order even at $\theta =0$ when $\theta$ is large. 
This result is important. 
If the chiral transition becomes first order at $\mu =0$, it will change the scenario of cosmological evolution. 
For example, the first-order transition allows us to think the inhomogeneous Big-Bang nucleosynthesis model or a new scenario of baryogenesis.
\acknowledgments
This work was supported by JSPS KAKENHI Grant Number 23-2790.
The numerical calculations were performed on the HITACHI SR16000 at Kyushu University 
and the NEC SX-9 at CMC, Osaka University.


\begin{thebibliography}{99}
\bibitem{Baker}
C. A. Baker {\it et al.}, Phys. Rev. Lett. {\bf 97}, 131801 (2006).
\bibitem{Kawarabayashi}
K. Kawarabayashi and N. Ohta, Nucl. Phys. B {\bf 175}, 477 (1980); 
Prog. Theor. Phys. {\bf 66}, 1789 (1981); 
N. Ohta, Prog. Theor. Phys. {\bf 66}, 1408 (1981); 
{\bf 67}, 993(E) (1982).
\bibitem{Negro}
M. D'Elia and F. Negro, Phys. Rev. Lett. {\bf 109}, 072001 (2012).
\bibitem{Sasaki-theta}
T. Sasaki, J. Takahashi, Y. Sakai, H. Kouno, and M. Yahiro, Phys. Rev. D {\bf 85}, 056009 (2012).
\bibitem{tHooft}
G. 't Hooft, Phys. Rev. Lett. {\bf 37}, 8 (1976); 
Phys. Rev. D {\bf 14}, 3432 (1976); 
{\bf 18}, 2199(E) (1978).
\bibitem{Kobayashi-Maskawa}
M. Kobayashi and T. Maskawa, Prog. Theor. Phys. {\bf 44}, 1422 (1970); 
M. Kobayashi, H. Kondo, and T. Maskawa, Prog. Theor. Phys. {\bf 45}, 1955 (1971).
\bibitem{Roessner}
S. R\"{o}\ss ner, C. Ratti, and W. Weise, Phys. Rev. D {\bf 75}, 034007 (2007).
\bibitem{Sakai-epnjl}
Y. Sakai, T. Sasaki, H. Kouno, and M. Yahiro, Phys. Rev. D {\bf 82}, 076003 (2010).
\bibitem{Sakai-Z3}
 Y. Sakai, K. Kashiwa, H. Kouno, and M. Yahiro, Phys. Rev. D {\bf 77}, 051901 (2008).
\bibitem{Kondo}
K.-I. Kondo, Phys. Rev. D {\bf 82}, 065024 (2010).
\bibitem{Braun}
J. Braun, L.M. Haas, F. Marhauser, and J.M. Pawlowski, Phys. Rev. Lett. {\bf 106}, 022002 (2011); 
J. Braun and A. Janot, Phys. Rev. D {\bf 84}, 114022 (2011).
\bibitem{Wetterich}
C. Wetterich, Phys. Lett. B {\bf 301}, 90 (1993).
\bibitem{Sasaki-Columbia}
T. Sasaki, Y. Sakai, H. Kouno, and M. Yahiro, Phys. Rev. D {\bf 84}, 091901 (2011).
\bibitem{RW}
A. Roberge and N. Weiss, Nucl. Phys. B {\bf 275}, 734 (1986).
\bibitem{Forcrand}
P. de Forcrand and O. Philipsen, Phys. Rev. Lett. {\bf 105}, 152001 (2010).
\bibitem{DElia}
M. D'Elia and F. Sanfilippo, Phys. Rev. D {\bf 80}, 111501 (2009).
\end{thebibliography}
\end{document}